# RECENT TRENDS AND RESEARCH ISSUES IN VIDEO ASSOCIATION MINING

Vijayakumar.V [1] and Nedunchezhian.R [2]

[1] Research Scholar, Bharathiar University, Coimbatore, &
Department of Computer Applications, Sri Ramakrishna Engineering College,
Coimbatore, Tamil Nadu, India - 641 022
veluvijay20@gmail.com

[2] Professor and Head, Department of Information Technology,
Sri Ramakrishna Engineering College, Coimbatore, Tamil Nadu, India-641 022
rajuchezhian@gmail.com

## ABSTRACT

*With the ever-growing digital libraries and video databases, it is increasingly important to understand and mine the knowledge from video database automatically. Discovering association rules between items in a large video database plays a considerable role in the video data mining research areas. Based on the research and development in the past years, application of association rule mining is growing in different domains such as surveillance, meetings, broadcast news, sports, archives, movies, medical data, as well as personal and online media collections. The purpose of this paper is to provide general framework of mining the association rules from video database. This article is also represents the research issues in video association mining followed by the recent trends.*

## 1. INTRODUCTION

Multimedia data is being acquired at an increasing rate due to technological advances in sensors, computing power, and storage. Multimedia Data Mining is the process of extracting previously unknown knowledge and detecting interesting patterns from a massive set of multimedia data [3]. Video is rapidly becoming one of the most popular multimedia due to its high information and entertainment capability. It also consists of audio, video and text together.

Video mining is a process which can not only automatically extract content and structure of video, features of moving objects, spatial or temporal correlations of those features, but also discover patterns of video structure, objects activities, video events, etc. from vast amounts of video data without little assumption about their contents [1][28]. Many video mining approaches have been proposed for extracting useful knowledge from video database. Finding desired information in a video clip or in a video database is still a difficult and laborious task due to its semantic gap between the low-level feature and high-level video semantic concepts. Video data mining can be classified in following categories, such as pattern detection, video clustering and classification and video association mining [2][8].

One of the important problems in video data mining is video association rule mining. Mining association rule from video data is usually a straightforward extension of association rule mining in transaction databases. Video Association Mining is a relatively new and emerging research trend. It is the process of discovering associations in a given video. It aims to extract interesting correlations, frequent patterns, associations or casual structures among sets of items in the video databases. This technique is an extension of data mining to image domain. It is an inter disciplinary field that combines techniques like computer vision, image processing, data mining, machine learning, data base and artificial intelligence. A lot of work was developed to find the association in the traditional transactional database. Video association mining is still in its infancy, and an under-explored field. Only limited work was developed in this area.

A video database contains lot of semantic information. The semantic information describes what is happening in the video and also what is perceived by human users. The semantic information of a video has two important aspects [4][8][10]. They are (a). A spatial aspect which means a semantic content presented by a video frame, such as the location, characters and objects displayed in the video frame. (b). A temporal aspect which means a semantic content presented by a sequence of video frames in time, such as character's action and object's movement presented in the sequence. To represent temporal aspects, the higher-level semantic information of video is extracted by examining the features audio, video, and superimposed text of the video. The semantic information includes the detecting trigger events, determining typical and anomalous patterns of activity, generating person-centric or object-centric views of an activity, classifying activities into named categories, and clustering and determining the interactions between entities. The temporal aspect of videos prevents the efficient browsing of these very large databases. Many efforts are conducted to extract the association between low-level visual features and high-level semantic concepts for image annotation [17].

The remainder of the paper is organized as follows. Section 2 introduces the basic concepts of video association rule mining. The general framework is presented in Section 3. Section 4 shows the recent trends in video association rule mining, and section 5 highlights a few researches issues and Section 6 concludes this paper.

## 2. BASIC CONCEPTS

Association rule discovery is an unsupervised data mining technique for discovering relationships among sets of variable values (items) from very large datasets. The first algorithm mining association rule was proposed by Agrawal, Imielinski, and Swami in 1993[9].

Given a set of transactions, where each transaction contains a set of items, an association rule is defined as an expression X ->Y , where X and Y are sets of items and $X \cap Y = null$ .The rule implies that the transactions of the database which contain X tend to contain Y . There are three measures of the association: support, confidence and interest. The support factor indicates the relative occurrence of both X and Y within the overall data set of transactions and is defined as the ratio of the number of tuples satisfying both X and Y over the total number of tuples. The confidence factor is the probability of Y given X and is defined as the ratio of the number of tuples satisfying both X and Y over the number of tuples satisfying X. an interest factor is defined to indicate the usefulness of the rules. The interest factor is a measure of human interest in the rule. For example, a high interest means that if a transaction contains X, then it is much more likely to have Y than the other items.

The association rule problem is to identify all association rules that satisfy a user specified minimum support and minimum confidence and this is solved in two steps. Firstly, all item sets whose support is greater than the given minimum are discovered and these are called frequent item sets. This step is computationally and I/O intensive. Given m items, there can be potentially $2^m$ item sets. Efficient methods are needed to traverse this exponential growth of itemsets search space to enumerate all the frequent itemsets. Second. Frequent itemsets are then used to generate interesting association rules where a rule is considered as interesting if its confidence is higher than the minimum confidence. This step is easier, but the overall performance of a mining algorithm is determined by the first step.

Video association mining differs from traditional association mining. Video data is governed by temporal properties while images are bound by spatial properties. Temporal information in a video sequence plays an important role in conveying video content. Temporal pattern mining differs from the traditional ARM in two aspects. First, an item set in traditional ARM contains only distinct items without considering the quantity of each item in the item set. Temporal pattern mining searches the whole video to identify the frequent item sets. In event detection, it is indispensable that an event is characterized by not only the attribute type but also its occurrence frequency. Second, in traditional ARM, the order of the items appeared in a transaction is considered as irrelevant. Therefore, transaction {a, b} is treated the same as {b, a} [2] [11]. Since changing the position of frequent items in a pattern leads to another pattern which is different in the video database. For instance, in video association mining the pattern ABCD differs from ABDC because in the second pattern D occurs before C.

Two types of association are identified in video data. First, Intra associations are those in which all items involved in the association are the same such as visually similar shots of the same object taken from different view points. For example in movie database the shot consist of similar objects in different view to impress the viewers. Second, Inter association are those which consist of items of different types, which are scenes that consist of visually distinct shots of different objects. For example, the surveillance video database consists of different objects in the different shots.

Amongst the above mentioned techniques, video inputs posses temporal properties, the frequent temporal pattern mining process is governed by two data specific parameters namely temporal support and temporal distance threshold. Frequent temporal pattern mining can also be treated as Sequence Pattern Mining. Sequential pattern mining finds frequently occurring patterns ordered by time. A sequence composed by a series of nominal symbols from a particular alphabet is usually called a temporal sequence and a sequence of continuous, real-valued elements, is known as a time series. Temporal data mining is to discover hidden relations between sequences and sub-sequences of events. The discovery of relations between sequences of events involves mainly three steps: the representation and modelling of the data sequence in a suitable form; the definition of similarity measures between sequences; and the application of models and representations to the actual mining problems.

Zhu, X., Wu, X., Elmagarmid, A., Feng, Z. and Wu, L. proposed the various definitions and evaluation measures for video associations by taking the distinct features of video data into consideration and proposed a solution in mining patterns from video stream that consists of multiple information sources (image, audio, and caption text) [2].

Video association as sequential pattern with $\{X_1,....X_i....X_j; X^t_i < X^t_j \text{ for any } i<j\}$ where $X_i$ is a video item L denotes the length of the association, $X_1, \cap....\cap X_i,....\cap X_j=\phi$, $X^t_i$ denotes the temporal order of $X^i$ and $X^t_i < X^t_j$ indicated that $X^t_i$ happens before $X^t_j$. The sequence patterns obtained from the video association mining differs with patterns derived by the conventional frequent pattern mining. Due to the fact that the temporal information in a video sequence plays an important role in conveying the video content, so traditional association measures {support and confidence} are integrated with video temporal information to evaluate video associations.

Shirahama, K.  Ideno, K. and  Uehara, K  represented the movie data as a multi-stream of raw level Meta data. Then, they proposed a video data mining approach with temporal constraint for extracting previously unpredictable semantic patterns from a movie, and also they proposed parallel pattern mining algorithm in order to reduce the expensive computation time [19].

## 3. GENERAL FRAMEWORK

Video consists of a series of images. An image is called a frame. A shot is defined as an unbroken sequence of frames taken from one camera. The video information can be accessed in two approaches. 1. Shot-based approach and 2. Object based approach. The shot based analysis is conducted at the shot-level, the adjacent shots are deemed as the transaction and the attributes (items) can be the feature descriptors (low-, mid- or object-level extracted from different channels) or event types in the transaction. An object-based representation for video data can facilitate video search and content analysis very effectively. Normally, this is done through spatiotemporal segmentation and region tracking.

There are two kinds of videos in our daily life: videos with some content structure and videos without any content structure. The movie and news videos are used to convey the video content. The surveillance, and sports videos does not consists any video content [4].  Depends upon the kinds of video database, the accessing approach will be chosen.

The major steps involved in the video association rule mining system are: pre-processing, feature extraction, database construction and association rule generation. In general, the video data must be first pre-processed to convert the unstructured raw data format in structured format. Subsequently, the video data undergo various transformations and features extraction to generate the important features from the video data and store the extracted data in structured databases. With the generated structured video database, mining can be carried out using data mining techniques to discover significant patterns. These resulting patterns are then evaluated and interpreted to obtain the knowledge of the application. The overall system framework is shown in Fig.1.

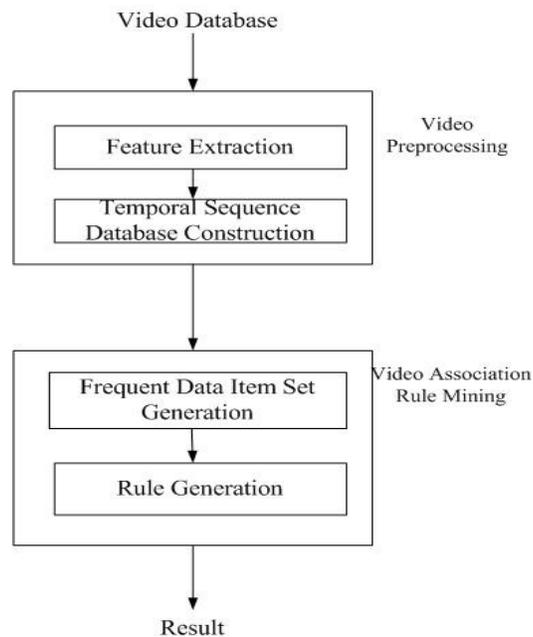

Figure 1. Video Association Mining Framework

## 3.1. Video Pre-processing

Generally, video data are unstructured data source. So, the knowledge can not be extracted directly. To convert in a structured format the video data is parsed into video shots. Discovering the shot boundary is the first of the pre-processing phase. A video shot can be considered as a basic unit of video data. Video data consists of audio, video and text together. Visual features are captured with the assistance of color analysis and object segmentation techniques; Audio features are exploited in both time-domain and frequency-domain and text features are detected using video text processing techniques. All the features are considered to temporally segment the raw video sequences into a set of consecutive video shots in the pre-processing process. Then key frames are extracted from the video shot applying the shot partitioning algorithm. After detecting the key-frames, multimodal features were extracted from the video source and in the second step it constructs video temporal data sequence.

### 3.1.1. Feature Extraction

Image processing algorithms are used to extract visual features from the key frame images. Audio signal classification systems are employed the extraction of a set of audio features from the input audio signal. Pitch, Timbrel features (Zero crossings, Mel frequency cepstral coefficients), Rhythm features (Beat strength, Rhythmic regularity), Low energy rate are different types features extracted from the audio signal. Text data present in video contain useful information for mining the associations. There are two types of video text are presented in the video. First, the text shown in video scenes (scene text), and the second is the text post-processed and added into the video, such as team names and their scores, referred as Superimposed text. Superimposed text in video sequences provides useful information about their contents. Once we detected a text change in the sequence, a symbolic tag is added at the corresponding place in the data sequence. All the features are organized into video hybrid sequence (feature vectors). The feature vectors are then employed to model them as transactions, which are then used in the mining process.

### 3.1.2. Video Data Construction

Generally, each and every symbolic stream (visual, audio, text and objects with frame window) is assigned a symbol for constructing the video sequence. Each key frame in the video is treated as a time unit and transform the extracted features of each time unit into symbolic streams according to the Look-up Table (consists the equivalent symbol for every feature) mapping. Finally, the original video data is transformed into temporal video sequence which consists of multiple streams into single stream. The transformed structured video sequence is used to mine the association in video database.

## 3.2. Video Association Mining

It aims to extract the interesting frequent patterns, associations among the set of items in a video temporal sequence database. It contains two steps. (i). Discovering the frequent sub sequence; (ii). Rule generation.

### 3.2.1. Frequent Subsequence Generation

Several algorithms have been proposed for discovering frequent temporal subsequence in video Sequence. Frequent temporal sequential pattern mining fall into two big categories: Apriori-like and FP-growth-like. The Apriori based approach is still limited by the high cost of multiple database scans and candidate generation. Apriori algorithm uses a level-wised and iterative approach; it first generates the candidates then test them to delete the non-frequent itemsets. Most of previous studies adopted an Apriori-like candidates generation-and-test approach [2] [13]. Comparing with Apriori algorithm, FP-growth algorithm has following features. (1) It uses FP-tree to store the main information of the database. The algorithm scans the database only twice, avoids multiple database scans and reduces I/O time. (2) It does not need to generate candidates, reduces the large amount of time that is consumed in candidates generation and test. (3) It uses a divide-and-conquer approach in the mining process, so the searching space is significantly decreased. The efficiency of the FPgrowth algorithm is about an order of magnitude faster than the Apriori algorithm [12][15].

B. SivaSelvan and N.P. Gopalan presented an m-ary tree based frequent temporal pattern mining algorithm eliminates the repeated scans limitation of Apriori based FTP mining algorithm  and requires only two overall original input scans. Frequent and Infrequent single patterns were identified by the first scan. Various possible patterns' temporal count was generated by the second scan, which updates the m-ary tree constructed at the end of first scan. A simple tree traversal subject to the minimum support threshold generated the complete frequent temporal patterns. This technique concentrates on reducing the overhead incurred as a result of the tree updating strategies [30].

The efficiency of frequent item-set mining algorithms is determined mainly by three factors: the way candidates are generated, the data structure that is used and the implementation details.

### 3.2.2.  Rule generation

The knowledge to be discovered is in the form of association rules, which are mined from temporal video sequence retrieved from a large corpus. Work on rule generations mainly focus on newer algorithms, deriving more types of association rules, and interestingness of the rules.

## 4. RECENT TRENDS IN VIDEO ASSOCIATION MINING

Video association rules are an important type of knowledge representation revealing implicit relationships among the items present in a large video sequence. It can be applied in the various fields of applications domains such as medical, surveillance, movie, sports and etc., for extracting high level video concepts, event detection, summary generation, classification.

### 4.1. Concept Detection

The objective of the semantic concept extraction is to capture information about how events in the extracted clip are related to one another. Mahesh Goyani, Shreyash Dutta, Gunvatsinh Gohil and Sapan Naik proposed an algorithm to detect semantic concepts from cricket video. The proposed scheme works in two phases. In first phase a top-down event detection and classification is performed using hierarchical tree. It consists of five levels.  key frames are identified based on Hue Histogram difference at level 1. At level 2, logo transitions classified the frames as real-time or replay. At level 3, the real time frames classified as field view, pitch view or non field view based on thresholds like Dominant Soli Pixel Ration (DSPR) and Dominant Grass Pixel Ration. At level 4, close up and crowd frames detected based upon edge detection. At level 5, the close up frames classified into player of team A, player of team B and umpire based upon skin colour and corresponding jersey colour. Then crowd frames classified into spectators, player's gathering of team A or player's gathering of team B. In the second stage, labels are associated with each frame event, which is used as input to A-Priori algorithm for concept mining [26].

Lin Lin, Guy Ravitz, Mei-Ling Shyu, and Shu-Ching Chen extracted multimodal audio and visual features from the broadcast and used the apriori algorithm. Association rule mining is to find frequent item-sets in the feature data set and generate classification rules to classify video shots to the different shots classifying the concepts such as weather, sports, and commercial. The association rule mining technique is adopted to bridge the semantic gap between low-level multi-modal features and the concepts of interest [10].

Association rule mining is used automatically enrich the representation of video content through a set of semantic concepts based on concept co-occurrence patterns. Maheshkumar H. Kolekar, Kannappan Palaniappan computed the association in the soccer video for the events of each excitement clip using a priori mining algorithm. They proposed a novel sequential association distance to classify the association of the excitement clip into semantic concepts. They considered goal scored by team-A, goal scored by team-B, goal saved by team-A, goal saved by team-B as semantic concepts [24].

## 4.2. Event detection

Video events contain rich semantic information. These are normally defined as the interesting events which capture user attentions. For example, a soccer goal event is defined as the ball passing over the goal line without touching the goal posts and the crossbar.

Zongxing Xie, Mei-Ling Shyu, and Shu-Ching Chen proposed a framework to achieve a fully automatic video event detection procedure via the combination of distance based and rule-based data mining techniques. The C4.5 decision tree was used for event detection as it is a well-know rule-based algorithm good at learning the associations among different features of a set of pre-labeled instances [22].

Min Chen, Shu-Ching Chen, and Mei-Ling Shyu presented a hierarchical temporal association mining approach integrated association rule mining and sequential pattern discovery to systematically determine the temporal patterns significant for characterizing the events and results in a set of temporal rules, which are effectively employed to capture the candidate video events [11].

Maheshkumar H. Kolekar, and  Kannappan Palaniappan proposed a hierarchical framework for soccer (football) video event sequence detection and classification [24].

## 4.3. Summary

X. Zhu and X. Wu proposed an association based video summarization scheme for news videos, medical videos and movie which consists video pre processing, association mining, and summary creation [20].

Maheshkumar H. Kolekar, and Kannappan Palaniappan extracted excitement clips with semantic concept label to summarize many hours of video to collection of soccer highlights such as goals, saves, corner kicks, etc. These results with correctly indexed soccer scenes, enabling structural and temporal analysis, such as video retrieval, highlight extraction, and video skimming [24].

## 4.3. Knowledge Discovery

Video Association mining provides more efficient support for video database management. Xingquan Zhu and Xindong Wu designed video association mining scheme to mine sequentially associated clusters from the video sequence to convey valuable knowledge for video database management [23].

Zhu, X., Wu, X., Elmagarmid, A., Feng, Z. and Wu, L. proposed multilevel video association mining to mine video knowledge with integrating video processing and existing data mining algorithms [2].

Chao Liao, Patricia P. Wang, and Yimin Zhang divided the MTV programs into music video clips. Then, video and audio features are extracted from the video clips. The dual-wing harmonium model combined these features and discovers association patterns by clustering the sample data in hidden semantic space. They generated similar-style MTV by inference most related video clip from raw video with respect to each music clip in a given song. The association patterns can be used to measure the similarity between the generated result and the original MTV. They also used the discovered patterns to facilitate automatic MTV generation [27].

B. SivaSelvan and N.P. Gopalan presented a video association mining scheme to generate association in video. It can be used to predict futuristic events based on the occurrence of a certain sequence of events frequently. The association employed in video classification to determine the overall nature of the movies such as being romantic, comic, etc. Video associations are also employed in summarization by including the most frequent patterns in the summary [16].

Movie database serve as good test case for mining different types of knowledge [18][19][21] such as movie highlight, interesting events (emotions, mood, serenity, violence level etc., crying indicates sadness or happiness, clapping or laughing indicates happiness, bomb blast or a gun shot indicates street violence), event pattern (violence is followed by sad scene, movies classification, frequent events (frequent events are every day human actions that frequently appear in the movie) and rhythms of character's topic (some kinds of rhythms associated with contents; rhythm of romantic scene is very slow and suspense scene is very fast).

## 4.4. Indexing

Zhu, X., Wu, X., Elmagarmid, A., Feng, Z. and Wu, L. proposed association rule mining technique to construct a sport video indexes. They extracted visual and audio features using video processing techniques and applied multilevel sequential association mining to explore associations, classified the associations by assigning each of them with a class label, and used their appearances in the video to construct video indices [2].

## 4.5. Classification

The generated associations can be employed to perform video classification. For example, we determine the nature of the movie and can classified the movies into different categories such as being romantic, tragedy, comedy, etc., using video association mining. Associative classification algorithm discovers the association rules with the frequency count (minimum support) and ranking threshold (minimum confidence) while restricted to the concepts (class labels). Classification using Association Rule Mining takes advantages of its high accuracy and ability to handle large databases. It is another major predictive analysis technique that aims to discover a small set of rule in the database that forms an accurate classifier [6]. Three main research aspects for associative classification have emerged. One is to improve the support and confidence measurements themselves. Another one is to use other evaluation criteria such as lift, coverage, leverage, and conviction. The last one is to use an integrated algorithm to generate association rules. Well-known Associative rule classification algorithms are decision tree classifier, support vector machine classifier, and traditional association rule classifier.

Lin Lin, Mei-Ling Shyu, Guy Ravitz and Shu-Ching Chen proposed a framework with a new associative classification algorithm which generates the classification rules based on the correlation between different feature-value pairs and the concept classes by using Multiple Correspondence Analysis in a video database [5].

Weighted Associative Classifiers is another concept that assigns different weights to different features and can get more accuracy in predictive modelling system [6]. Different weights can be assigned to different attributes according to their predicting capability. The temporal aspect can be used to improve the prediction capability.

Lin Lin, and Mei-Ling Shyu, proposed a video semantic concept detection framework was built upon a new weighted association rule mining (WARM) classifier algorithm. The WARM classifier algorithm was able to capture the different significance degrees of the items (feature-value pairs) in generating the association rules for video semantic concept detection. The framework first applied multiple correspondence analyses (MCA) to project the features and classes into a new principle component space and discovered the correlation between feature-value pairs and classes. Next, it considered both correlation and percentage information as the measurement to weight the feature-value pairs and to generate the association rules. Finally, it performed classification by using these weighted association rules [7].

N. Harikrishna, Sanjeev Satheesh, S. Dinesh Sriram, and K.S. Easwarakumar proposed method to segment a cricket video into shots and identified the visual content in them. Using sequential pattern mining and support vector machine, they classified the sequence of shots into four events, namely RUN, FOUR, SIX and OUT. The cricket video was then summarized based on user-supplied parameters [25].

Ling Chen, Sourav S. Bhowmick and LiangTien Chia presented a video classifier which employs the association rule mining technique to discover the actual dependence relationship between video states. The discriminatory state transition patterns mined from different video categories are then used to perform classification. Besides capturing the association between states in the time space, they captured the association between low-level features in spatial dimension to further distinguish the semantics of videos [29].

## 5. RESEARCH ISSUES AND FUTURE DIRECTIONS

In this section, we have examined several major research issues and challenges and raise in video association mining, and pointed out some promising research directions.

### 5.1. Data Processing Model

Video data models are used to define and concisely describing the structural video properties. The data models for representing video have to be general and broad enough to accommodate the range of formats, types of video data (scripted or unscripted) and lengths of different types of programs, whether the video is a two hour movie, a one-hour talk-show, a five-minute home video clip, or a thirty-second news segment. Nowadays, various video processing techniques are available to explore visual and audio cues for association mining and it will inevitably incur information loss from the original video sequences to transferred symbolic streams [2]. More studies are needed to address this issue in the mining activities. The issue of data processing model here is to find an efficient representation way to extract the video symbolic sequence for association rule mining from the overall video stream and the temporal order must be maintained in the sequence.

### 5.2. Algorithm issues

Another fundamental issue is to choose the right type of mining algorithms. Mostly, the mining algorithms are mainly derived from the existing data mining schemes (with some extensions for video mining scenarios); so, extensive studies are needed to explore efficient mining algorithms which are unique for mining knowledge from video data.

There were also studies needed to improve the speed of finding large itemsets with hash table, map, and tree data structures [14]. Compare to the numerous works done to search for better algorithms to mine large itemsets in a video sequence database, the qualifying criterion – support threshold – and the mechanism behind it – counting, temporal distance between the items – have received much less attention. The problem with support threshold is that it requires expert knowledge, which is subjective at best, to set this parameter in the system. Setting it arbitrarily low will qualify itemsets that should be left out, vice versa. Moreover, as database size increases, the support threshold may need to be adjusted.

Discovering association rules has been studied over a decade. Research is also done in the types of association rules. At the beginning of data mining research, Boolean association rules dominated the field. Later on, more focus was put on Quantitative Association Rules. A more thrust was focused on the performance and scalability of the algorithms, but not enough attention was given to the quality (interestingness) of the rules generated.

## 5.3. Application Domains

The general framework was adopted in the various applications such as, sports and news. The performance of the classifiers can be improved with additional events such in cricket video as Wide, LBW, etc.

The current framework can be extended to many domains such as movies, sports, education and evaluate the performance of the video mining algorithm in environments with more events. The most promising domain is the surveillance video, where the routine vehicles in security areas normally comply with some associations like enter ->stop -> drop off -> leave and a vehicle which does not comply with this association might be problematic and deserves further investigation. However, due to the inherent differences between different video domains (e.g., the concept of shot and video text do not exist in surveillance videos), need more activities to analyze the video content details for association mining, e.g., extract trails and status of moving objects to characterize associations [2]. In future, the trend will be to turn the attention to the application of these researches in various areas of our lives, e.g. educational, medicine, homeland security and surveillance, etc.

## 5.4. Memory issues

The next fundamental issue we need to consider is how to optimize the memory space consumed when running the mining algorithm. This includes how to collect the information from video database and how to choose a compact in-memory data structure that allows the video information to be stored, updated and retrieved efficiently. Fully addressing these issues in the mining algorithm can greatly improve its performance. This is due to bounded memory size and multi-dimensional representation of video database.

## 5.5. Performance and Scalability issues

Only limited work was done to improve the performance and scalability of video association mining. The computational cost of association rules mining can be reduced by reducing the number of passes over the video database, adding extra constraints on the structure of patterns and parallelization. There is need for faster and more scalable algorithms because data in video databases are getting bigger day by day.

The list of issues associated with the video association mining is shown in Table. 1.

Table 1. Issues of Video Association Mining

| Task | Approaches / Applications | Issues |
|------|---------------------------|--------|
| Preprocessing | Shot level, Frame level, Region (object ) level, scene level | ➢ Depends on video data model and application domain |
| Feature Extraction and Transformation | Visual features , audio features, text features, motion features | ➢ Visual features and audio features are sensitive to the parameters (number of bin, selection of parameters, selection of boundaries); ➢ Motion calculation computationally expensive; ➢ Multi models feature presentation needs domain knowledge, synchronized, computation, reliable. |

| Video association mining | Apriori based association find sequence of events | <ul><li>Scalability (number of patterns generated );</li><li>Efficient data structure is need;</li><li>Finding semantic boundaries;</li><li>Setting minimum temporal support and temporal distance threshold is data dependent;</li><li>Importance of individual items in the sequence;</li></ul> |
|---|---|---|

## 6. CONCLUSIONS

Discovery association rules in the video database will be a more thrust area in future. In this paper, we have examined an important research directions and research issues in video association mining. And also, we have presented an efficient general framework for mining the association rules in the video data base. Generally, the video association mining facing lot of challenges such as, bridging semantic gap between the low-level features and high-level semantic concepts, No well-defined data models for representing video contest, Synchronizing with multiple feature streams(audio, video, text), data transformation form video database to structured database, High dimensionality and temporal aspect of video database.

Future work includes improving the feature extraction step to obtain more representative features, efficient frequent pattern searching method for the multi-stream, extracting complex events in different types of video domains, unique efficient mining algorithms for mining knowledge from video data and newer rule types may be necessary to facilitate new data analysis.

The motivation of doing video association mining is to use the discovered patterns to improve decision making. It has therefore attracted significant research efforts in developing algorithms to extract the useful knowledge and do domain specific tasks for data from domains such as surveillance, meetings, broadcast news, sports, movies, medical data, as well as personal and online media collections.

## REFERENCES

[1]   Bhatt, C.A. &  Kankanhalli, M.S., (2011) "Multimedia data mining: state of the art and challenges", *Multimedia Tools Appl.,* Vol. 51,  pp. 35–76.

[2]   Zhu, X., Wu, X., Elmagarmid, A., Feng, Z. & Wu, L., (2005) "Video Data Mining: Semantic Indexing and Event Detection from the Association perspective", *IEEE Transactions on Knowledge and Data Engineering,* Vol. 17, No.5, pp. 1-14.

[3]   Oh, J., Lee, J., & Hwang, S., (2005) "Video Data Mining," Idea group.

[4]   Marcela X. Ribeiro, Agma J. M. Traina, Caetano Traina, Jr., & Paulo M. Azevedo-Marques, (2008) "An Association Rule-Based Method to Support Medical Image Diagnosis With Efficiency", *IEEE Transactions On Multimedia,* Vol. 10, No. 2,  pp. 277-285.

[5]   Lin Lin, Mei-Ling Shyu, Guy Ravitz & Shu-Ching Chen, (2009) "Video Semantic Concept Detection via Associative Classification", *In the proceedings of IEEE International Conference on Multimedia and Expo (ICME09), USA,* pp. 418-421.

[6]   Sunita Soni  & O.P.Vyas, (2010) "Using Associative Classifiers for Predictive Analysis in Health Care Data Mining", *International Journal of Computer Applications ,* Vol 4 – No.5, pp. 33-37.

[7]   Lin Lin, , & Mei-Ling Shyu, (2010) "Weighted association rule Mining for Video Semantic Detection", *International Journal of Multimedia Data Engineering and Management,* Vol 1- No-1, pp. 37-54.

[8]  Fatemi, Nastaran, Poulin, Florian, Raileany, Laura E. & Smeaton, Alan F, (2009) "Using association rule mining to enrich semantic concepts for video retrieval", *In proceedings of International Conference on Knowledge Discovery and Information Retrieval*, Portugal.

[9]  Agrawal R, Imielinski T, & Swami A , (2009) "Mining association rules between sets of items in large databases", *In Proceedings of ACM SIGMOD Conference on Management of Data, Washington DC, USA,* pp 207-216.

[10]  Lin Lin, Guy Ravitz, Mei-Ling Shyu,  & Shu-Ching Chen, (2007) "Video Semantic Concept Discovery Using Multimodal-Based Association Classification", *In Proceedings of ICME*, pp. 859-863.

[11]  Min Chen, Shu-Ching Chen, & Mei-Ling Shyu, (2007) "Hierarchical Temporal Association Mining for Video Event Detection in Video Databases", *In Proceedings of IEEE 23rd International Conference on Data Engineering Workshop.*

[12]  J. Han, J. Pei,  & Y. Yin, (2000) "Mining frequent patterns without candidate generation",  *In Proceedings of  ACM-SIGMOD Int'l Conf on Management of Data (SIGMOD'00). Dallas, TX, New York: ACM Press,* pp. 1-12.

[13]  R. Agrawal, & R. Srikant, (1994) "Fast algorithms for mining association rules", *In Proceedings of the 20th Int'l Conf on Very Large Databases (VLDB'94), Santiago*, pp. 487-499.

[14]  Nan Jiang & Le Gruenwald, (2006) "Research Issues in Data Stream Association Rule Mining", *SIGMOD Record*, Vol. 35, No. 1, pp.14-20.

[15]  Jian Pei (2002) "Pattern-Growth Methods for Frequent Pattern Mining", *Doctoral Dissertation* Simon Fraser University Burnaby, BC, Canada, Canada.

[16]   B. SivaSelvan & N.P. Gopalan, (2007) "Efficient algorithms for video association mining" *In Proceedings of 20th Conference of the Canadian Society for Computational Studies of Intelligence, Canada*, pp. 250-260.

[17]  Ishwar  K. Sethi, Ioana Coman, and Daniela Stan, (2001) "Mining Association Rules between Low-level Image Features and High-level Concepts," *In Proceedings of SPIE: Data Mining and Knowledge Discovery III,* pp.279-290.

[18]  Y. Matsuo, K. Shirahama, & K. Uehara,(2003) "Video Data Mining: Extracting Cinematic Rules from Movie", *In Proceedings of Int'l Workshop Multimedia Data Management (MDM-KDD).*

[19]  Shirahama, K.,  Ideno, K., &  Uehara, K, (2005) "Video data mining : mining semantic patterns with temporal constraints from movies", *In Proceedings of Seventh IEEE International Symposium on Multimedia.*

[20]  X. Zhu & X. Wu, (2003) "Sequential Association Mining for Video Summarization," *In Proceedings of IEEE Int'l Conf. Multimedia and Expo,* vol. 3, pp. 333-336.

[21]  D. Wijesekera & D. Barbara,(2000) "Mining Cinematic Knowledge: Work in Progress," *In Proceedings Int'l Workshop Multimedia Data Management (MDM-KDD).*

[22]  Zongxing Xie, Mei-Ling Shyu, & Shu-Ching Chen (2007) "Video Event Detection With Combined Distance-Based And Rule-Based Techniques", *In Proceedings of IEEE International Conference on Data Mining Multimedia and Expo,* pp. 2026-2030.

[23]  Xingquan Zhu & Xindong Wu (2003) "Mining Video Associations for Efficient Database Management",  *In Proceedings of the 18th international joint conference on Artificial intelligence,* Morgan Kaufmann Publishers Inc. USA , pp.1422-1425.

[24]  Maheshkumar H. Kolekar, & Kannappan Palaniappan (2009) "Semantic Concept Mining Based on Hierarchical Event Detection for Soccer Video Indexing", *Journal Of Multimedia*, Academy Publisher, Vol. 4, No. 5, pp. 298-313.

[25]  N. Harikrishna, Sanjeev Satheesh, S. Dinesh Sriram, K.S. Easwarakumar (2011) "Temporal Classification of Events in Cricket Videos ", *In proceedings of National Conference on Communications (NCC),* pp. 1 – 5.

[26] Mahesh Goyani, Shreyash Dutta, Gunvatsinh Gohil & Sapan Naik (2011) " Wicket Fall Concept Mining From Cricket Video Using A-Priori Algorithm", *The International Journal of Multimedia & Its Applications (IJMA)* Vol.3, No.1, pp. 111-121.

[27] Chao Liao, Patricia. Wang, & Yimin Zhang, (2009) "Mining Association Patterns between Music and Video Clips in Professional MTV" *In Proceedings of MMM*, Springer-Verlag Berlin Heidelberg, pp. 401–412.

[28] Dong-Jun Lan, Yu-Fei Ma, & Hong-Jiang Zhang (2003) " A novel motion-based representation for video mining", *In Proceedings of the 2003 International Conference on Multimedia and Expo* - Vol 03.

[29] Ling Chen, Sourav S. Bhowmick & LiangTien Chia, (2004) " VRules: An Effective Association based Classifier for Videos", *In Proceedings of the 2nd ACM international workshop on Multimedia databases*, ACM New York, NY, USA.

[30] B. Gopalan N.P & SivaSelvan B,(2006) "An m-ary tree based Frequent Temporal Pattern Mining Algorithm". *In Proceedings of 6th IEEE Indicon International Conference on Emerging Trends in Information and Communication Technology*, New Delhi.

**Authors**

**Vijayakumar.V**, Assistant Professor in the Department of Computer Applications, Sri Ramakrishna Engineering College, Coimbatore, Tamil Nadu, India. He is currently pursuing his doctoral degree at Bharathiar University, Coimbatore, Tamil Nadu, India in the area of Multimedia Data Mining. He obtained his M.C.A degree and M.Phil degree in Computer-Science from Bharathiar University, Coimbatore. His research interests are Data Mining, Multimedia Information Retrieval, Image and Video Processing. He has presented four papers in National / International Conference and one in journal. He has guided several undergraduate, post-graduate projects and seven M.Phil research scholars. He is a student member of IEEE and life member of ISTE.

**Dr. Nedunchezhian. R.** is working as the Professor and Head in the department of Information Technology, Sri Ramakrishna Engineering College, Coimbatore. He has more than 18 years of experience in research and teaching. Currently, he is guiding many Ph.D scholars of the Anna University, Coimbatore, and the Bharathiar University. His research interests are knowledge discovery and data mining, distributed computing, and database security. He has published many research papers in national/international conferences and journals. He has edited a book entitled "Handbook of Research on Soft Computing Applications for Database Technologies: Techniques and Issues" which was published by IGI publications, USA in April, 2010. He is a Life member of Advanced Computing and Communication Society and ISTE.